\documentclass[conference]{IEEEtran}
\IEEEoverridecommandlockouts
\usepackage{cite}
\usepackage{amsmath,amssymb,amsfonts}
\usepackage{algorithmic}
\usepackage{graphicx}
\usepackage{textcomp}
\usepackage{tikz}
\usepackage{pgfplots}
\usepackage{xcolor}
\usepackage{bbm}
\usepackage{soul,color}
\usepackage{makecell}
\usepackage[font=small,skip=2pt]{caption}
\usepackage{adjustbox}
\usepackage{hyperref}
\usepackage{array}

\usepackage[acronyms,nomain,xindy,nowarn]{glossaries}
\makeglossaries
\newacronym{5g}{5G}{Fifth-Generation Wireless Systems}
\newacronym{ai}{AI}{artificial intelligence}
\newacronym{ann}{ANN}{Artificial Neural Network}
\newacronym{ask}{ASK}{amplitude-shift keying}
\newacronym[plural={AWGN}, \glsshortpluralkey={AWGN}]{awgn}{AWGN}{additive white Gaussian noise}
\newacronym{au}{AU}{aerial user}
\newacronym{ap}{AP}{access point}
\newacronym{bec}{BEC}{binary erasure channel}
\newacronym{ber}{BER}{bit error rate}
\newacronym{bler}{BLER}{block error rate}
\newacronym{bpsk}{BPSK}{binary phase-shift keying}
\newacronym{bs}{BS}{base station}
\newacronym{bbu}{BBU}{baseband unit}

\newacronym{cdf}{CDF}{cumulative distribution function}
\newacronym{cnn}{CNN}{convolutional neural network}
\newacronym{csi}{CSI}{channel state information}
\newacronym{comp}{CoMP}{coordinated multi-point}
\newacronym{cran}{C-RAN}{cloud-radio access network}
\newacronym{cca}{CCA}{cloud coverage area}

\newacronym{dnn}{DNN}{deep neural network}
\newacronym{drl}{DRL}{deep reinforcement learning}
\newacronym{dqn}{DQN}{deep Q-network}
\newacronym{ecc}{ECC}{error-correcting code}
\newacronym{ecdf}{ECDF}{empirical distribution function}
\newacronym{ee}{EE}{energy efficiency}
\newacronym{ec}{EC}{edge cloud}
\newacronym{cc}{CC}{central cloud}

\newacronym{ff}{FF}{feed-forward}
\newacronym{ffnn}{FF-NN}{feed-forward neural network}
\newacronym{fsk}{FSK}{frequency-shift keying}
\newacronym{fs}{FS}{functional split}

\newacronym{gm}{GM}{Gaussian mixture}
\newacronym{GOPS}{GOPS}{Giga Operation Per Second}

\newacronym{hmappo}{H-MAPPO}{hierarchical multi-agent proximal policy optimization}
\newacronym{hdrl}{HDRL}{Hierarchical deep reinforcement learning}
\newacronym{hmarl}{HMARL}{Hierarchical Multi-Agent Reinforcement Learning}

\newacronym{iid}{i.i.d.}{independent and identically distributed}
\newacronym{il}{IL}{information leakage}
\newacronym{iot}{IoT}{internet of things}

\newacronym{lb}{LB}{lower bound}
\newacronym{los}{LoS}{line-of-sight}

\newacronym{mac}{MAC}{multiple access channel}
\newacronym{map}{MAP}{maximum a posteriori}
\newacronym{mc}{MC}{Monte Carlo}
\newacronym{mdp}{MDP}{Markov decision process}
\newacronym{mimo}{MIMO}{multiple-input multiple-output}
\newacronym{miso}{MISO}{multiple-input single-output}
\newacronym{ml}{ML}{machine learning}
\newacronym{mmwave}{mmWave}{millimeter wave}
\newacronym{mop}{MOP}{multi-objective programming problem}
\newacronym{mrc}{MRC}{maximum ratio combining}
\newacronym{msc}{MSC}{modulation and coding scheme}
\newacronym{mse}{MSE}{mean squared error}
\newacronym{msac}{MSAC}{masked soft actor-critic}
\newacronym{mappo}{MAPPO}{multi-agent proximal policy optimization}
\newacronym{madrl}{MADRL}{multi-agent deep reinforment learning}
\newacronym{milp}{MILP}{Mixed Integer Linear Programming}

\newacronym{nlos}{NLoS}{non-line-of-sight}
\newacronym{nn}{NN}{neural network}
\newacronym{nve}{NVE}{normalized validation error}
\newacronym{ric}{Near-RT RIC}{Near-Real Time RAN Intelligent Controller}
\newacronym{noma}{NOMA}{Non-orthogonal multiple access}

\newacronym{oru}{O-RU}{O-RAN radio unit}
\newacronym{oran}{O-RAN}{open-radio access network}
\newacronym{odu}{O-DU}{O-RAN distributed unit}

\newacronym{pdf}{PDF}{probability density function}
\newacronym{pwc}{PWC}{polar wiretap code}
\newacronym{ppo}{PPO}{proximal policy optimization}
\newacronym{pf}{PF}{processing function}
\newacronym{qam}{QAM}{quadrature amplitude modulation}
\newacronym{qos}{QoS}{quality of service}

\newacronym{ra}{RA}{rearrangement algorithm}
\newacronym{rbf}{RBF}{radial basis function}
\newacronym{relu}{ReLU}{rectified linear unit}
\newacronym{ris}{RIS}{reconfigurable intelligent surface}
\newacronym{rl}{RL}{reinforcement learning}
\newacronym{rnn}{RNN}{recurrent neural network}

\newacronym{sac}{SAC}{soft actor-critic}
\newacronym{sgd}{SGD}{stochastic gradient descent}
\newacronym{sgp}{SGP}{secure goodput}
\newacronym{simo}{SIMO}{single-input multiple-output}
\newacronym{sinr}{SINR}{signal-to-interference-plus-noise ratio}
\newacronym{siso}{SISO}{single-input single-output}
\newacronym{snr}{SNR}{signal-to-noise ratio}
\newacronym{svm}{SVM}{support vector machine}

\newacronym{tvd}{TVD}{total variation distance}

\newacronym{uav}{UAV}{unmanned aerial vehicle}
\newacronym{ub}{UB}{upper bound}
\newacronym{urllc}{URLLC}{ultra-reliable low latency communication}

\newacronym{v2v}{V2V}{vehicle-to-vehicle}
\newacronym{v2x}{V2X}{vehicle-to-everything}

\newacronym{zoc}{ZOC}{zero-outage capacity}
\setacronymstyle{long-short}
\glsdisablehyper
\newcommand{\todo}[2][]{\ignorespaces
	\if\relax\detokenize{#1}\relax
	{\color{red}[TODO: #2]}%
	\else
	{\color{red}[TODO (#1): #2]}%
	\fi
}

\usepackage{physics}

\def\BibTeX{{\rm B\kern-.05em{\sc i\kern-.025em b}\kern-.08em
    T\kern-.1667em\lower.7ex\hbox{E}\kern-.125emX}}

\begin{document}

\title{Hierarchical Multi Agent DRL for Soft Handovers Between Edge Clouds in Open RAN
}

\setlength{\textfloatsep}{0.3pt}

\author{%
\IEEEauthorblockN{%
Federico Giarrè\IEEEauthorrefmark{1}, Irshad A. Meer\IEEEauthorrefmark{1}, Meysam Masoudi\IEEEauthorrefmark{2}, Mustafa~Ozger\IEEEauthorrefmark{3}\IEEEauthorrefmark{1}, and  Cicek Cavdar\IEEEauthorrefmark{1}}
\IEEEauthorblockA{\IEEEauthorrefmark{1}Division of Communication Systems, KTH Royal Institute of Technology, Sweden}
\IEEEauthorblockA{\IEEEauthorrefmark{3}Department of Electronic Systems, Aalborg University, Denmark}
\IEEEauthorblockA{\IEEEauthorrefmark{2}Ericsson AB, Sweden}

Email: fgiarre@kth.se, iameer@kth.se, meysam.masoudi@ericsson.com, mozger@es.aau.dk, cavdar@kth.se}

\maketitle

\begin{abstract}

Multi-connectivity (MC) for aerial users via a set of ground access points offers the potential for highly reliable communication. Within an open radio access network (O-RAN) architecture, edge clouds (ECs) enable MC with low latency for users within their coverage area. However, ensuring seamless service continuity for transitional users—those moving between the coverage areas of neighboring ECs—poses challenges due to centralized processing demands. To address this, we formulate a problem facilitating soft handovers between ECs, ensuring seamless transitions while maintaining service continuity for all users. We propose a hierarchical multi-agent reinforcement learning (HMARL) algorithm to dynamically determine the optimal functional split configuration for transitional and non- transitional users. Simulation results show that the proposed approach outperforms the conventional functional split in terms of the percentage of users maintaining service continuity, with at most 4\% optimality gap. Additionally, HMARL achieves better scalability compared to the static baselines.

\end{abstract}

\begin{IEEEkeywords}
Functional Split, Handover, O-RAN, Hierarchical MARL\end{IEEEkeywords}

\section{Introduction}\label{introduction}
\begin{figure}
    \centering
    \includegraphics[width=0.7\linewidth]{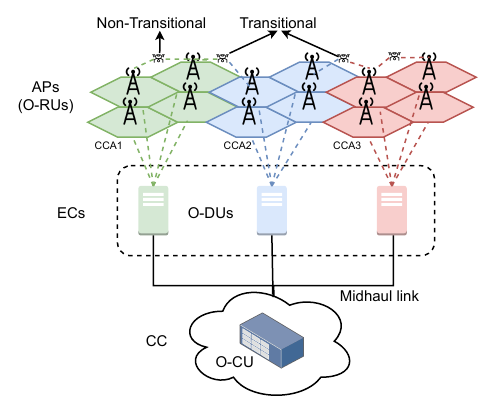}
    \caption{Overview of the considered \gls{oran} architecture.}
    \label{fig:hcran}
\end{figure}

Multi-connectivity (MC) and functional split (FS) are key features for \gls{oran} architecture in 6G networks \cite{mahmoud20216g,rihan2023ran}.
While MC leverages the distributed nature of radio access to achieve stringent 6G \gls{qos} requirements, FS allows for flexible management of resources between the central and edge nodes of the network~\cite{larsen_survey_2018}. In MC, each user is concurrently connected to and served by a cluster of distributed \glspl{ap}. To support this connectivity, coordination is centralized at a shared location accessible to all \glspl{ap} within the serving cluster~\cite{bassoy_coordinated_2017}.
 The \gls{oran} architecture facilitates scalable multi-connectivity by disaggregating the RAN protocol stack into modular components, enabling flexible deployment and enhanced coordination across network nodes~\cite{oran}.
Specifically, the \gls{oran} Distributed Unit (O-DU), acting as an \gls{ec}, coordinates distributed \glspl{ap} (O-RU) to form serving clusters for users within a \gls{cca}, managed by the \gls{oran} Centralized Unit (O-CU), as shown in~\autoref{fig:hcran}.

\begin{figure}
    \centering
    \includegraphics[width=\linewidth]{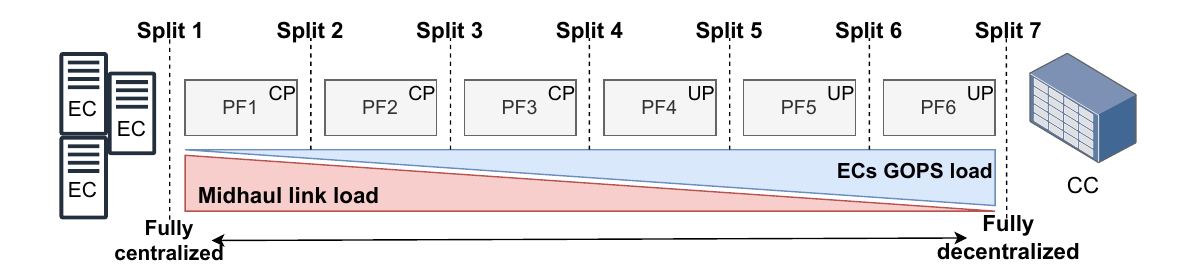}
    \caption{Functional splits and resulting tradeoffs in terms of resources.}
    \label{fig:split}
\end{figure}

During user mobility, seamless service continuity is achieved through dynamic reconfiguration of the serving clusters, where new \glspl{ap} are added, and some are dropped. This process, referred to as soft handover, ensures uninterrupted service throughout the user’s transition.
Centralized processing for soft handovers is achieved by distributing \glspl{pf} for users through the FS paradigm, enabling FSs between the \gls{ec} and \gls{cc}.
This approach allows flexibility in deciding where each function is executed, as illustrated in \autoref{fig:split}.

Mobility management for MC users within a single \gls{cca} is handled through dynamic reconfiguration of serving clusters at each \gls{ec}~\cite{Meer2024, meer2024hierarchical}. However, managing MC users transitioning between \glspl{cca}, termed as transitional users, remains a significant challenge, particularly under stringent \gls{qos} requirements. Transitional users include \glspl{ap} from multiple \glspl{cca} in their serving cluster, as seen in~\autoref{fig:hcran}, necessitating a shift in user coordination to the nearest common node between the \glspl{cca}, i.e., the \gls{cc}.

A fixed FS deployment is commonly used but presents significant challenges in balancing the needs of transitional and non-transitional users (those moving within a single \gls{cca}). Optimizing FS for transitional users is essential to enable soft handovers and maintain service continuity during transitions between \glspl{cca}. However, this optimization often increases centralization, leading to higher network-wide delays that can degrade the performance of non-transitional users. Conversely, prioritizing non-transitional users limits the MC capabilities required by transitional users, risking interruptions in their service continuity. In addition, as illustrated in \autoref{fig:split}, different FS configurations impose varying resource demands in terms of necessary bandwidth in the midhaul link and computational resources at \glspl{ec}. This complicates the decision-making process in resource-constrained networks.

This paper addresses these challenges by proposing a soft handover mechanism within the \gls{oran} architecture. The mechanism ensures seamless service continuity for transitional users, adheres to stringent \gls{qos} requirements, and minimizes disruptions for non-transitional users by dynamically adapting to resource constraints through optimal FS configurations across various network levels.
The main contribution of this paper are given as follows:
\begin{itemize}
    \item We formulate a mixed-integer linear programming (MILP) problem to ensure service continuity by identifying flexible FS configurations that support soft handovers, and meet user stringent QoS requirements using limited network resources.
   
    \item We propose a \gls{hmarl} algorithm to solve the optimization problem, where a high-level agent manages soft handovers for transitional users and low-level agents optimize the performance of non-transitional users within their \gls{cca}.
   
    \item We introduce a novel communication mechanism within \gls{hmarl}, where the high-level agent and low-level agents interact in a turn-based manner, simulating the message-passing process of a distributed system.
    \item We evaluate our proposed \gls{hmarl} algorithm against industry-standard FS configurations and the optimal solution derived through brute force to assess its effectiveness. 
\end{itemize}


\section{Related Works}\label{rw}

The problem of soft handovers between \glspl{ec} remains unaddressed in existing literature, though related sub-problems like MC, flexible FS, and soft handovers have gained interest. For instance, \cite{wang_handover_2016} introduces MC for soft handovers using a greedy clustering algorithm, reducing handovers but failing to support MC users transitioning between clusters.

In \cite{alba_dynamic_2022}, a 5G/6G RAN design dynamically adjusts FS configurations based on network conditions, balancing adaptation benefits and performance costs but without considering MC. 
The work in \cite{matoussi_user-centric_2021} employs multiple flexible FS configurations tailored to specific user classes based on the service requirements. A \gls{drl} algorithm determines near-optimal FS configurations for each user class. Although effective in resource allocation, this study does not address specific service requirements such as delay or service continuity. Similarly, \cite{alabbasi_delay-aware_2017} presents a detailed delay model for centralized RAN and its relationship to FS, considering the computational weight of each \gls{pf} relative to a reference architecture. However, the study does not account for dynamic environments involving user mobility.

To the best of our knowledge, no prior work addressed the integration of flexible FS and MC within \gls{oran} networks to facilitate soft handovers for users roaming both within a single \gls{cca} (intra-CCA) and across multiple \glspl{cca} (inter-CCA).


\section{System Model}\label{modelling}

We define a set of aerial users as $\mathcal{U}$, with $\mathcal{T} \subset \mathcal{U}$ representing transitional users. The network is composed of a set of \glspl{ap} $\mathcal{B}$, grouped into \glspl{cca}. Each \gls{ec} $e \in \mathcal{E}$ is responsible for the processing of \glspl{pf} for the users and \glspl{ap} in a \gls{cca}. Let $\mathcal{B}_u^t$ denote the serving cluster of \glspl{ap} for user $u$ at time $t$ with $\abs{\mathcal{B}_u^t} \geq 2$. Furthermore, $\mathcal{U}^{e,t}$ denotes the set of users connected to \gls{ec} $e$ at time $t$.

We use a greedy user-centric clustering algorithm for MC, which forms clusters with \glspl{ap} nearest to the users with a minimum cluster size of two. 
A user is a transitional user, if the the serving cluster $\mathcal{B}_u$ consists of \glspl{ap} from different CCAs as shown in Figure~\ref{fig:hcran}.
The transitional status of a user $u$ is formalized as follows:
\begin{equation}
    {transitional}(u)=
    \begin{cases}
      True, \quad \text{if} \quad u \in \{\mathcal{U}^{e} \cap \mathcal{U}^{e'}\} \\
      False, \quad \text{otherwise}
    \end{cases}.
\end{equation}

We define the \gls{pf} chain as comprising six \glspl{pf}, as illustrated in \autoref{fig:split}, aligning with FS research \cite{alabbasi_delay-aware_2017,wang_interplay_2017,noauthor_small_nodate,masoudi_energy-optimal_2022} and 3GPP FS options~\cite{larsen_survey_2018}. The \glspl{pf} are: (\textit{PF1}) radio signal processing, baseband conversion, and serial-to-parallel conversion; (\textit{PF2}) cyclic prefix removal and FFT; (\textit{PF3}) resource demapping and cell processing; (\textit{PF4}) equalization, inverse FFT, QAM, and multi-antenna processing; (\textit{PF5}) error correction, HARQ, and turbo decoding; and (\textit{PF6}) higher-layer functions. \glspl{pf} $1$–$3$ are cell-specific, while \glspl{pf} $4$–$6$ are user-specific \cite{alabbasi_delay-aware_2017}.  

For the remainder of this paper, we adopt the FS notation in the rightmost column of \autoref{tab:splits}. \glspl{ap} in the MC cluster of non-transitional users leverage their common \gls{ec} as a centralization point, enabling MC without FS constraints. However, for transitional users, \glspl{pf} related to MC coordination and scheduling must be deployed in the \gls{cc}, limiting FS options for MC support. While additional 3GPP FS options can facilitate MC for transitional users, they introduce extra constraints \cite{larsen_survey_2018}. Thus, this paper considers only FS $1$–$4$ as enablers for transitional users' MC.

         
         
    

\begin{table}[t]
    \centering
    \caption{Comparison of FS Notations in the Literature.}
    \label{tab:splits}
    \begin{tabular}{l|l|l}
    \hline
    \textbf{3GPP FS Options \cite{larsen_survey_2018}} & \textbf{Wang \emph{et al}. \cite{noauthor_small_nodate}} & \textbf{Our FS Options \cite{alabbasi_delay-aware_2017}} \\
    \hline
    FS 1     & RRC-PDCP         & FS 7     \\
    FS 2     & PDCP-RLC         & FS 6     \\
    FS 3     & RLC              & -        \\
    FS 4     & RLC-MAC          & -        \\
    FS 5     & MAC              & -        \\
    FS 6     & MAC-PHY          & FS 5     \\
    FS 7-1   & PHY              & FS 4     \\
    FS 7-2   & PHY              & FS 3     \\
    FS 7-3   & PHY              & FS 2     \\
    FS 8     & PHY-RF           & FS 1     \\
    \hline
    \end{tabular}
\end{table}

\begin{figure}
    \centering
    \includegraphics[width=0.8\linewidth]{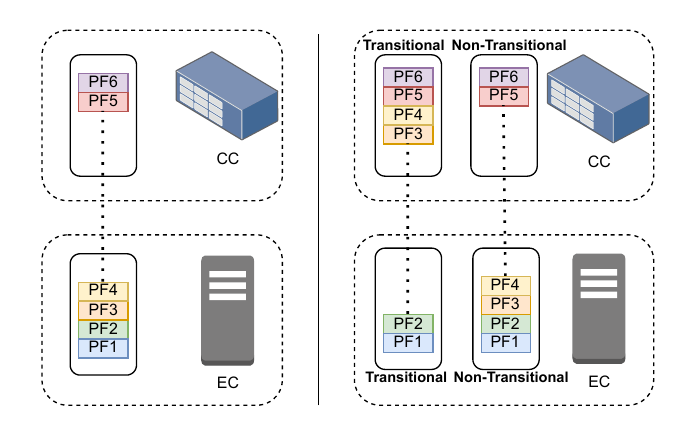}
    \caption{Traditional functional split configuration (left), User-grouped functional split configuration (right).}
    \label{fig:multichain}
\end{figure}

Users are categorized based on their status, i.e., transitional or non-transitional as illustrated in \autoref{fig:multichain}, with a distinct FS configuration applied to each group. This grouping enables the deployment of one FS configuration to facilitate soft handovers for transitional users and another to minimize delays for non-transitional users. Each \gls{ec} handles \glspl{pf} for both transitional and non-transitional users within its assigned \gls{cca}.
For simplicity, in the remainder of this paper, the FS configuration for a user group $x$ is called $F_x$, with $x\in X= \{\text{transitional, non-transitional}\}$.

Each \gls{pf} has specific computational demands that impact network resources. The \gls{GOPS} requirements at \glspl{ec} and the midhaul data rate for each FS option are derived from~\cite{wang_centralize_2017}. 

Let $G_b$ and $G_u$ represent the required \glspl{GOPS} for deploying Cell \glspl{pf} and User \glspl{pf}, respectively. Additionally, let $\mathcal{U}_{x,e}$ and $\mathcal{B}^e$ denote the set of users in group $x$ and the \glspl{ap} connected to \gls{ec} $e$. The \gls{GOPS} required at \gls{ec} $e$ at time $t$ for user group $x$ is given by: \vspace{-3mm}
\begin{subequations}
 \begin{equation}\label{GOPS}
    G^{e,t}_x = \sum^{\mathcal{B}^e}_{b} G_b(F_x^t) + \sum^{\mathcal{U}_{x,e}}_{u} G_u(F_x^t).
     \vspace{-1mm}
\end{equation}
The total amount of GOPS required at time $t$ at EC $e$ can then be computed as:      \vspace{-3mm}
\begin{equation}
    G^{e,t}_{tot} = \sum^X_x G^{e,t}_x.
\end{equation}
\end{subequations}

The required midhaul bandwidth scales not with the number of users connected to the \gls{cca}, but with the number of \glspl{ap} in the \gls{cca}. The total midhaul resources needed by \gls{ec} $e$ at time $t$ for user group $x$ is computed as:\vspace{-2mm}
   \begin{equation}
    M^{e,t}_x = \sum^{\mathcal{B}^e}_{b} M(F_x^t),
\end{equation}
where $M(F_x^t)$ is the requested midhaul resources for $F_x^t$ given in \cite{wang_centralize_2017}. 
Since the midhaul link is shared between ECs, we calculate the total required resources as:
\begin{equation}
\vspace{-2mm}
    M_{tot}^t = \sum^\mathcal{E}_e \sum^X_x M^{e,t}_x.
\end{equation}
A service $k$, imposes per user constraints on the maximum acceptable end-to-end (E2E) latency, $D_{th}^k$, and the maximum acceptable outage, $\varepsilon^k$. 
 
\emph{Service reliability} of a user $u$ is defined as the probability of the user experiencing an E2E delay lower than the delay threshold for the deployed service $k$ \cite{salehi_reliability_2023}:
\begin{equation}\label{reliability}
    \rho_u = 1-\underbrace{P(D_u > D^k_{th})}_{\varepsilon_u}, 
\end{equation}
where $D_u$ is experienced E2E delay perceived by user $u$. The delay is computed by taking into account the time for transmission of data, the processing time of \glspl{pf}, and the hardware overhead of the network as in \cite{alabbasi_delay-aware_2017}.

A user $u$ achieves service continuity if $\varepsilon_u \leq \varepsilon_k$. Service interruptions, such as disconnections or user drops, can compromise reliability and service continuity. A user is considered disconnected under two conditions: \textit{i)} undergoing a complete cluster reconfiguration, or \textit{ii)} experiencing an signal to noise plus interference ratio (SINR) below a threshold ${\textit{SINR}}_{th}$. If the network lacks sufficient resources to process a user's \glspl{pf} under the current FS configuration, the user is dropped.

\section{Problem Formulation}
Our objective is to ensure service continuity for users while managing a limited set of available resources. Hence, given available resources and users' information such as position and status at timestep $t$, we select suitable FS configuration for each user group.
For a generic set of users $\Omega$, the ratio of users achieving service continuity is calculated as:
\begin{equation}
    R(\Omega)=\frac{1}{|\Omega|}\sum^U_u \mathbf{1}(\varepsilon_u < \varepsilon_k).
\end{equation}
Denoting $\mathcal{F} = \{F_x, \forall x \in X\}$ as the set of all FS configurations, we aim to maximize the rate of users achieving service continuity:  \vspace{-3mm} \begin{subequations}
    \begin{equation}\label{obj}
    \max\limits_{\mathcal{F}} \quad  \omega_{nt}R(\mathcal{U}\setminus\mathcal{T}) +\omega_{t} R(\mathcal{T})\quad 
\end{equation}
\begin{equation}\label{c1}
    \text{s.t.} \quad G^{e,t}_{tot} \leq G_{th}, \quad \forall e \in \mathcal{E} \; ,\forall t
\end{equation}
\begin{equation}\label{c2}
     M^{t}_{tot} \leq M_{th}, \quad \forall t,
\end{equation}
with $\mathcal{U}\setminus\mathcal{T}$ the set of non-transitional users. \eqref{obj} is composed of two parts: $R(\mathcal{U}\setminus\mathcal{T})$ is the ratio of non-transitional users achieving service continuity, and $R(\mathcal{T})$ is the same ratio but for transitional users. The objective is the weighted sum of them, where $\omega_t$ and $\omega_{nt}$ are used for transitional and non-transitional users, respectively. Network resources are limited; thus, at any given time, the consumption of \gls{GOPS} and midhaul bandwidth is constrained by the thresholds $G_{th}$ and $M_{th}$, respectively, as defined in \eqref{c1} and \eqref{c2}. 
Due to the linear objective, constraints, discrete decision space, and continuous variables, the  formulated problem is a MILP problem. However, it is shown in \cite{alba_dynamic_2022} that in dynamic scenarios, the optimal FS may change every few seconds, rendering optimal solvers impractical at \gls{oran}'s scale. Given the dynamic nature of mobility and resource availability at each timestep, an ML-based solution is better suited to address the formulated problem. 

\end{subequations}

\section{Proposed HMARL Approach}

To address the optimization problem, we propose an \gls{hmarl} algorithm to manage decision-making for FS configurations, aligning with the hierarchical structure of the \gls{oran} architecture shown in~\autoref{fig:hcran}. HMARL is an RL approach in which high-level agents, based on the defined hierarchy, can influence or dictate the actions of lower-level agents. 
HMARL was selected as the solution approach for two main reasons: \textit{i)} Due to MC constraints, the FS configuration for transitional users has to be the same throughout the network. For this reason, we need a centralized agent that enforces such FS configuration for all CCAs. \textit{ii)} On the other hand, centralizing intra-CCA decisions undermines the benefits of O-RAN's heterogeneous, distributed architecture. 

\begin{figure}
    \centering
        \includegraphics[width=0.8\linewidth]{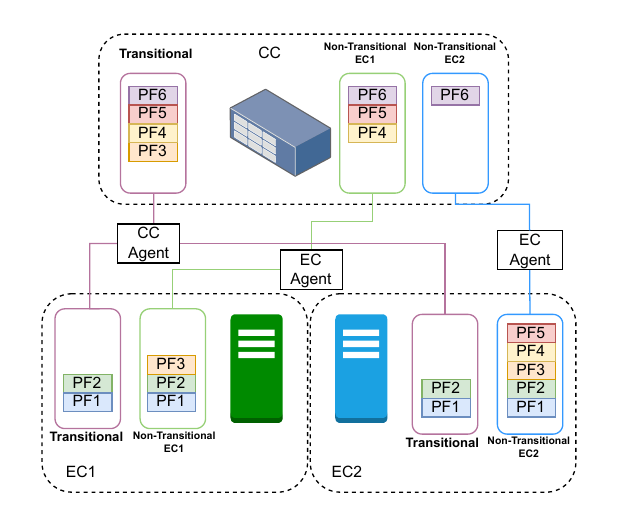}
    \caption{A CC agent enforces a common FS $F_t$, while ECs deploy individual FS $F_{nt}$.}
    \label{fig:hmarl}
\end{figure}

To simulate realistic agent communication, we incorporate turn-based messaging in \gls{hmarl}, mimicking the message-passing process of a distributed system. Due to distributed nature of the system, it would be unreasonable for every agent to have full knowledge about the network. For this reason, every timestep is divided in two turns: one for a high-level agent, and one for the low-level agents.  
A high-level agent operates in the \gls{cc}, acting first to enforce an FS configuration $F_t$ across the network. This decision is based on the number of transitional users, their service requirements, and their location. The high-level agent updates low-level agents on resource availability and the FS chosen for transitional users by adjusting their observations before their turn begins.
Low-level agents at each \gls{ec}, at the start of their turn, assess their respective CCAs and the additional information from the high-level agent, subsequently selecting the optimal FS configuration $F_{nt}$ to meet the requirements of non-transitional users using the remaining resources. An example of the resulting FSs in the network can be seen in \autoref{fig:hmarl}.


We present the dynamics of the proposed HMARL as a multi-agent cooperative stochastic game~\cite{busoniu_multi-agent_2010}. Given $\mathcal{N}$ the set of agents, this particular stochastic game can be described by the tuple
$   \langle S_{CC},\{S_{e}\}^\mathcal{E}_e,A_{CC},\{A_{e}\}^\mathcal{E}_e,\phi,r_{CC},    \{r_{e}\}^\mathcal{E}_e\rangle. $
Here, $S_i$, $A_i$ and $r_i$ represent the state space, action space and reward of an agent $i\in\{CC,\forall e\in\mathcal{E}\}$.  $\phi$ is the transition probability function to reach a new environment state after the joint action $A = A_{CC} \times \prod^\mathcal{E}_e A_e$ is applied to the current state. Each agent yields a specialized state space observation, action and reward based on their role.
%
\subsection{Low-level agents}
Low-level agents, placed on each \gls{ec}, are meant to deploy an FS configuration capable of serving non-transitional users in their \gls{cca} such that: \textit{i)} Users experience a low-enough delay to comply with their service requirements and maintain service continuity, and \textit{ii)} The FS deployed does not exceed the amount of resources available. To achieve such objectives, agents should have a thorough understanding of the resource consumption in the network and the number of users connected. For this reason, the state space for an agent at \gls{ec} $e$ at time $t$ can be defined by the tuple:
\begin{equation}\notag
    {s_{e}=(|\mathcal{U}^{e,t}|, D^{k,t}_{min}, E[|\mathcal{B}^e|], G^{e,t}_{tot},M^{e,t},M^{t}_{tot},a^\tau_{CC})}.
\end{equation}
Here, $|\mathcal{U}^{e,t}|$ denotes the number of users connected to its \gls{cca}, $D^{k,t}_{min}$ is the lowest delay limit to comply with, $E[|\mathcal{B}^e|]$ represents the mean number of users connected to \glspl{ap} in the CCA, $a^\tau_{CC}$ is the high-level agent action, and $G^{e,t}_{tot},M^{e,t},M^{t}_{tot}$ are the resources used in the network. The action space for a low-level agent $e$, $A_e$, is defined by the set of all FS, as there are no FS requirements for non-transitional users' MC. Finally, the reward given to these agents is calculated as:
\begin{equation}
    r_{e} = \frac{1}{|\mathcal{U}^e|} \sum_{u}^{\mathcal{U}^e}\mathbf{I}(D_u < D^k_{th}) - (\#dropped^e*\omega_{dc}),
\end{equation}
where we want connected users to comply with their delay requirements. A penalty is applied to the agents based on the number of dropped users. 
We assume all low-level agents learn a shared policy while receiving individual rewards. We also assume that CC's location can be exploited as parameter server to aggregate the individually learned weights of EC agents. This reduces the signaling overhead that would be generated by the otherwise fully distributed policy sharing. Furthermore, inference for the optimal FS configuration is necessary on a timescale of seconds \cite{alba_dynamic_2022}. Given this large timescale, we consider the exchanges between low and high-level agents, which consist of network observations and the high-level agent's actions, to be negligible.
%
\subsection{High-level agent}
The high-level agent, placed in the \gls{cc}, is in charge of imposing a FS configuration throughout the network for the handling of transitional users. Since the agent has virtually all the network resources available at any time by acting first, and our focus is to support transitional users, the state space observation of the agent at a time $t$ is formulated as the following tuple:
\begin{equation}
    {S_{CC}=(|\mathcal{T}|,D^{k,t}_{min},E[|\mathcal{B}_{\mathcal{T}}|])},
\end{equation}
where $|\mathcal{T}|$ is the amount of transitional users to be handled, $D^{k,t}_{min}$ is the lowest delay limit to comply with and $E[|\mathcal{B}_\mathcal{T}|]$  is the mean amount of users connected to \glspl{ap} serving transitional users.
While the higher-level agent yields a simpler state space observation, the action space $A_{CC}$ is restricted to the FS that can enable MC for transitional users. Finally, we model the reward for the higher-level agent as follows: \begin{equation}\label{rcc}
    r_{CC} = R(\mathcal{T}) + \sum^\mathcal{E}_e r_e.
\end{equation}

The agent is rewarded for maintaining the transitional users' service continuity during their soft-handover between CCAs ($R(\mathcal{T})$) while also receiving feedback on how its decision affects the rest of the agents (rightmost part of \eqref{rcc}).

\section{Numerical Evaluation}\label{impl}
\subsection{Deployment Scenario}


\begin{table}[t]
    \centering
    \caption{Training Parameters}
    \renewcommand{\arraystretch}{1.3} 
    \begin{tabular}{l|l|c}
    \hline
        \textbf{Symbol} & \textbf{Parameter Description} & \textbf{Value} \\
        \hline
        $|\mathcal{U}|$ & Total number of users & $50$ \\ 
        $|\mathcal{E}|$ & Number of ECs & $2$ \\
        $D^k_{\text{th}}$ & Delay limit for services & $12$ ms \\
        $\varepsilon^k$ & Maximum acceptable outage ratio & $10^{-5}$ \\
        $G_{\text{th}}$ & Computational limit at ECs & $16$ kGOPS \cite{alabbasi_delay-aware_2017,wang_centralize_2017} \\
        $M_{\text{th}}$ & Midhaul bandwidth limit & $30$ k$\times |\mathcal{E}|$ \\
        $|t|$ & Simulation duration & $300$ timesteps \\
        \hline
    \end{tabular}
    \label{tab:param}
\end{table}

To study the performance of the proposed HMARL algorithm, we implement a discrete time simulator in Python, complying with the Gym framework \cite{brockman2016openai}, and consider a standard library for the training and inference of RL agents. To train the agents for HMARL, the RLlib library \cite{liang2018rllib} has been used.  
\autoref{tab:param} summarizes the  simulation parameters. 
For the HMARL algorithm, two policies are trained using the Proximal Policy Optimization (PPO) \cite{schulman_proximal_2017}: one for the high-level agent at \gls{cc}, and one shared low-level policy for all agents at \glspl{ec}. PPO is chosen for its incremental policy updates, which support cooperative agent interactions and improve exploration and convergence.
The high-level policy uses a discount factor $\gamma = 0.99$, while the low-level policy uses $\gamma = 0.80$. This allows the agent placed at the \gls{cc} to be reactive in accommodating transitional users on a timestep basis, while the agents at \glspl{ec} can learn a more stable policy. Finally, in order to encourage the exploration of the environment, both policies are set up to have an entropy coefficient of $0.01$. 

\begin{figure}[b!]
    \centering
\begin{tikzpicture}

\definecolor{color0}{rgb}{0.67843137254902,0.847058823529412,0.901960784313726}

\begin{axis}[
legend cell align={left},
legend entries={GOPS load violations,
                Midhaul load violations},
legend style={
  fill opacity=0.8,
  draw opacity=1,
  text opacity=1,
  draw=white!80!black,
  at={(0.55,0.95)},
},
tick align=outside,
tick pos=left,
x grid style={white!69.0196078431373!black},
xmajorgrids,
xmin=0.5, xmax=11.5,
xtick style={color=black},
xtick={1.5,4.5,7.5,10.5},
xticklabels={Optimal,HMARL,F=3,F=4},
y grid style={white!69.0196078431373!black},
ylabel={Violation per episode ratio},
ymajorgrids,
ymin=-3.56666666666667, ymax=80,
ytick style={color=black},
ytick={0,10,20,30,40,50,60,70,80},
yticklabels={0,10,20,30,40,50,60,70,80},
height=0.5\linewidth,
width=\linewidth
]
\addlegendimage{only marks, mark=square*,color=color0,fill}
\addlegendimage{only marks, mark=square* ,color=red,fill}
\addplot [black, forget plot]
table {%
1 7.75
1 0
};
\addplot [black, forget plot]
table {%
1 20.0833333333333
1 33.6666666666667
};
\addplot [black, forget plot]
table {%
0.85 0
1.15 0
};
\addplot [black, forget plot]
table {%
0.85 33.6666666666667
1.15 33.6666666666667
};
\addplot [black, forget plot]
table {%
2 0
2 0
};
\addplot [black, forget plot]
table {%
2 0
2 0
};
\addplot [black, forget plot]
table {%
1.85 0
2.15 0
};
\addplot [black, forget plot]
table {%
1.85 0
2.15 0
};
\addplot [black, forget plot]
table {%
4 9.79166666666667
4 1.16666666666667
};
\addplot [black, forget plot]
table {%
4 22.625
4 36.5
};
\addplot [black, forget plot]
table {%
3.85 1.16666666666667
4.15 1.16666666666667
};
\addplot [black, forget plot]
table {%
3.85 36.5
4.15 36.5
};
\addplot [black, forget plot]
table {%
5 0
5 0
};
\addplot [black, forget plot]
table {%
5 1.41666666666667
5 3.33333333333333
};
\addplot [black, forget plot]
table {%
4.85 0
5.15 0
};
\addplot [black, forget plot]
table {%
4.85 3.33333333333333
5.15 3.33333333333333
};
\addplot [black, forget plot]
table {%
7 25.75
7 4.66666666666667
};
\addplot [black, forget plot]
table {%
7 47.1666666666667
7 66.1666666666667
};
\addplot [black, forget plot]
table {%
6.85 4.66666666666667
7.15 4.66666666666667
};
\addplot [black, forget plot]
table {%
6.85 66.1666666666667
7.15 66.1666666666667
};
\addplot [black, forget plot]
table {%
8 0
8 0
};
\addplot [black, forget plot]
table {%
8 0
8 0
};
\addplot [black, forget plot]
table {%
7.85 0
8.15 0
};
\addplot [black, forget plot]
table {%
7.85 0
8.15 0
};
\addplot [black, forget plot]
table {%
10 25.875
10 5.33333333333333
};
\addplot [black, forget plot]
table {%
10 46.5833333333333
10 71.3333333333333
};
\addplot [black, forget plot]
table {%
9.85 5.33333333333333
10.15 5.33333333333333
};
\addplot [black, forget plot]
table {%
9.85 71.3333333333333
10.15 71.3333333333333
};
\addplot [black, forget plot]
table {%
11 0
11 0
};
\addplot [black, forget plot]
table {%
11 0
11 0
};
\addplot [black, forget plot]
table {%
10.85 0
11.15 0
};
\addplot [black, forget plot]
table {%
10.85 0
11.15 0
};
\path [draw=color0, fill=color0]
(axis cs:0.7,7.75)
--(axis cs:1.3,7.75)
--(axis cs:1.3,20.0833333333333)
--(axis cs:0.7,20.0833333333333)
--(axis cs:0.7,7.75)
--cycle;
\path [draw=red, fill=red]
(axis cs:1.7,0)
--(axis cs:2.3,0)
--(axis cs:2.3,0)
--(axis cs:1.7,0)
--(axis cs:1.7,0)
--cycle;
\path [draw=color0, fill=color0]
(axis cs:3.7,9.79166666666667)
--(axis cs:4.3,9.79166666666667)
--(axis cs:4.3,22.625)
--(axis cs:3.7,22.625)
--(axis cs:3.7,9.79166666666667)
--cycle;
\path [draw=red, fill=red]
(axis cs:4.7,0)
--(axis cs:5.3,0)
--(axis cs:5.3,1.41666666666667)
--(axis cs:4.7,1.41666666666667)
--(axis cs:4.7,0)
--cycle;
\path [draw=color0, fill=color0]
(axis cs:6.7,25.75)
--(axis cs:7.3,25.75)
--(axis cs:7.3,47.1666666666667)
--(axis cs:6.7,47.1666666666667)
--(axis cs:6.7,25.75)
--cycle;
\path [draw=red, fill=red]
(axis cs:7.7,0)
--(axis cs:8.3,0)
--(axis cs:8.3,0)
--(axis cs:7.7,0)
--(axis cs:7.7,0)
--cycle;
\path [draw=color0, fill=color0]
(axis cs:9.7,25.875)
--(axis cs:10.3,25.875)
--(axis cs:10.3,46.5833333333333)
--(axis cs:9.7,46.5833333333333)
--(axis cs:9.7,25.875)
--cycle;
\path [draw=red, fill=red]
(axis cs:10.7,0)
--(axis cs:11.3,0)
--(axis cs:11.3,0)
--(axis cs:10.7,0)
--(axis cs:10.7,0)
--cycle;
\addplot [black, forget plot]
table {%
0.7 12.1666666666667
1.3 12.1666666666667
};
\addplot [black, forget plot]
table {%
1.7 0
2.3 0
};
\addplot [black, forget plot]
table {%
3.7 14.6666666666667
4.3 14.6666666666667
};
\addplot [black, forget plot]
table {%
4.7 0.333333333333333
5.3 0.333333333333333
};
\addplot [black, forget plot]
table {%
6.7 36.1666666666667
7.3 36.1666666666667
};
\addplot [black, forget plot]
table {%
7.7 0
8.3 0
};
\addplot [black, forget plot]
table {%
9.7 36.3333333333333
10.3 36.3333333333333
};
\addplot [black, forget plot]
table {%
10.7 0
11.3 0
};
\end{axis}

\end{tikzpicture}
    \caption{Constraint violation ratio per episode.}
    \label{fig:viol}
\end{figure}
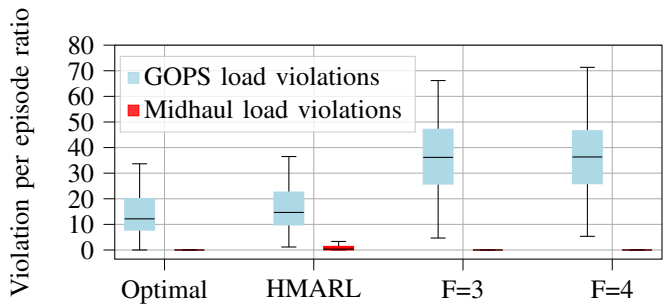

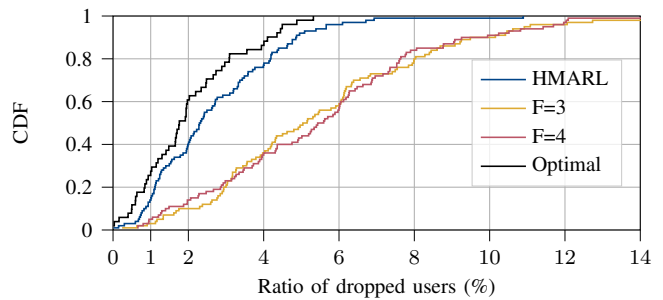
\begin{figure}
    \centering
\begin{tikzpicture}

\definecolor{color0}{HTML}{004488}
\definecolor{color1}{HTML}{DDAA33}
\definecolor{color2}{HTML}{BB5566}
\definecolor{color3}{HTML}{000000}
\definecolor{color4}{HTML}{AAAAAA}

\begin{axis}[
legend cell align={left},
legend style={fill opacity=0.8, draw opacity=1, text opacity=1, draw=white!80!black,at={(0.68,0.5)},anchor=west},
tick align=outside,
tick pos=left,
x grid style={white!69.0196078431373!black},
xlabel={Ratio of dropped users (\%)},
xmajorgrids,
xmin=0, xmax=14,
xtick={0,1,2,4,6,8,10,12,14,16,18},
xtick style={color=black},
y grid style={white!69.0196078431373!black},
ylabel={CDF},
ymajorgrids,
ymin=0, ymax=1,
ytick style={color=black},
width=.97\linewidth,
height=0.5\linewidth,
ytick={0,0.2,0.4,0.6,0.8,1},
yticklabels={0,0.2,0.4,0.6,0.8,1},
font={\footnotesize},
]
\addplot [semithick, color0, const plot mark left]
table {%
-inf 0
0.0133333333333333 0.01
0.146666666666667 0.02
0.3 0.03
0.593333333333333 0.04
0.673333333333333 0.05
0.7 0.06
0.726666666666667 0.07
0.746666666666667 0.08
0.773333333333333 0.09
0.82 0.1
0.873333333333333 0.11
0.933333333333333 0.12
0.933333333333333 0.13
0.98 0.14
1.00666666666667 0.15
1.02 0.16
1.04666666666667 0.17
1.08666666666667 0.18
1.08666666666667 0.19
1.11333333333333 0.2
1.12 0.21
1.13333333333333 0.22
1.14666666666667 0.23
1.20666666666667 0.24
1.24 0.25
1.24 0.26
1.26666666666667 0.27
1.28666666666667 0.28
1.34 0.29
1.4 0.3
1.49333333333333 0.31
1.53333333333333 0.32
1.56 0.33
1.62666666666667 0.34
1.78666666666667 0.35
1.86 0.36
1.94666666666667 0.37
1.95333333333333 0.38
1.97333333333333 0.39
1.98666666666667 0.4
2.00666666666667 0.41
2.03333333333333 0.42
2.05333333333333 0.43
2.07333333333333 0.44
2.13333333333333 0.45
2.15333333333333 0.46
2.22 0.47
2.28 0.48
2.28 0.49
2.29333333333333 0.5
2.32666666666667 0.51
2.35333333333333 0.52
2.4 0.53
2.42 0.54
2.42666666666667 0.55
2.49333333333333 0.56
2.62 0.57
2.64 0.58
2.7 0.59
2.71333333333333 0.6
2.72 0.61
2.76666666666667 0.62
2.91333333333333 0.63
3.1 0.64
3.19333333333333 0.65
3.2 0.66
3.3 0.67
3.32666666666667 0.68
3.32666666666667 0.69
3.36666666666667 0.7
3.43333333333333 0.71
3.48666666666667 0.72
3.58 0.73
3.58666666666667 0.74
3.70666666666667 0.75
3.76666666666667 0.76
3.99333333333333 0.77
4.00666666666667 0.78
4.14 0.79
4.17333333333333 0.8
4.19333333333333 0.81
4.19333333333333 0.82
4.22 0.83
4.36 0.84
4.39333333333333 0.85
4.62 0.86
4.63333333333333 0.87
4.76 0.88
4.76 0.89
4.85333333333333 0.9
4.87333333333333 0.91
4.95333333333333 0.92
5.08666666666667 0.93
5.38666666666667 0.94
5.66 0.95
5.66 0.96
6.09333333333333 0.97
6.71333333333333 0.98
6.93333333333333 0.99
10.8933333333333 1
};
\addlegendentry{HMARL}
\addplot [semithick, color1, const plot mark left]
table {%
-inf 0
0.24 0.01
0.633333333333333 0.02
0.906666666666667 0.03
1.12666666666667 0.04
1.16666666666667 0.05
1.33333333333333 0.06
1.33333333333333 0.07
1.61333333333333 0.08
1.67333333333333 0.09
1.74666666666667 0.1
2.3 0.11
2.37333333333333 0.12
2.59333333333333 0.13
2.62666666666667 0.14
2.76 0.15
2.78666666666667 0.16
2.84666666666667 0.17
2.9 0.18
2.93333333333333 0.19
2.94 0.2
3 0.21
3.03333333333333 0.22
3.04 0.23
3.13333333333333 0.24
3.16666666666667 0.25
3.17333333333333 0.26
3.17333333333333 0.27
3.26 0.28
3.27333333333333 0.29
3.44666666666667 0.3
3.54666666666667 0.31
3.56 0.32
3.68 0.33
3.79333333333333 0.34
3.91333333333333 0.35
3.97333333333333 0.36
4.02666666666667 0.37
4.12 0.38
4.18 0.39
4.20666666666667 0.4
4.22666666666667 0.41
4.30666666666667 0.42
4.31333333333333 0.43
4.34 0.44
4.5 0.45
4.62 0.46
4.8 0.47
4.92 0.48
4.96666666666667 0.49
5.04 0.5
5.15333333333333 0.51
5.16666666666667 0.52
5.30666666666667 0.53
5.36 0.54
5.45333333333333 0.55
5.48 0.56
5.78666666666667 0.57
5.91333333333333 0.58
6 0.59
6.04666666666667 0.6
6.06 0.61
6.1 0.62
6.12666666666667 0.63
6.13333333333333 0.64
6.14666666666667 0.65
6.17333333333333 0.66
6.19333333333333 0.67
6.32666666666667 0.68
6.34 0.69
6.38666666666667 0.7
6.56 0.71
6.78 0.72
6.83333333333333 0.73
7.41333333333333 0.74
7.49333333333333 0.75
7.54666666666667 0.76
7.84666666666667 0.77
7.94666666666667 0.78
7.98666666666667 0.79
7.99333333333333 0.8
8.03333333333333 0.81
8.28 0.82
8.4 0.83
8.44666666666667 0.84
8.60666666666667 0.85
8.68666666666667 0.86
9.08 0.87
9.21333333333333 0.88
9.24666666666667 0.89
9.48666666666667 0.9
10.1533333333333 0.91
10.42 0.92
10.46 0.93
10.6133333333333 0.94
10.92 0.95
11.08 0.96
12.06 0.97
12.7333333333333 0.98
15.62 0.99
17.3733333333333 1
};
\addlegendentry{F=3}
\addplot [semithick, color2, const plot mark left]
table {%
-inf 0
0.646666666666667 0.01
0.66 0.02
0.8 0.03
0.946666666666667 0.04
0.953333333333333 0.05
1.04666666666667 0.06
1.20666666666667 0.07
1.26666666666667 0.08
1.29333333333333 0.09
1.39333333333333 0.1
1.48666666666667 0.11
1.87333333333333 0.12
1.97333333333333 0.13
1.97333333333333 0.14
2.06 0.15
2.26 0.16
2.28 0.17
2.47333333333333 0.18
2.66666666666667 0.19
2.85333333333333 0.2
2.88 0.21
2.92666666666667 0.22
2.98 0.23
3.14666666666667 0.24
3.22666666666667 0.25
3.26666666666667 0.26
3.34666666666667 0.27
3.45333333333333 0.28
3.48 0.29
3.69333333333333 0.3
3.72666666666667 0.31
3.86 0.32
3.89333333333333 0.33
3.94666666666667 0.34
3.96 0.35
4 0.36
4.30666666666667 0.37
4.32666666666667 0.38
4.36 0.39
4.36 0.4
4.75333333333333 0.41
4.90666666666667 0.42
4.92666666666667 0.43
5.01333333333333 0.44
5.19333333333333 0.45
5.24 0.46
5.26666666666667 0.47
5.34 0.48
5.4 0.49
5.43333333333333 0.5
5.59333333333333 0.51
5.62 0.52
5.7 0.53
5.84 0.54
5.89333333333333 0.55
5.97333333333333 0.56
6 0.57
6.01333333333333 0.58
6.02 0.59
6.04 0.6
6.12 0.61
6.18666666666667 0.62
6.24666666666667 0.63
6.26666666666667 0.64
6.27333333333333 0.65
6.42666666666667 0.66
6.48 0.67
6.63333333333333 0.68
6.85333333333333 0.69
6.87333333333333 0.7
6.88666666666667 0.71
6.96 0.72
7.16666666666667 0.73
7.30666666666667 0.74
7.34666666666667 0.75
7.35333333333333 0.76
7.53333333333333 0.77
7.53333333333333 0.78
7.59333333333333 0.79
7.60666666666667 0.8
7.64 0.81
7.76 0.82
7.78666666666667 0.83
7.9 0.84
8.06 0.85
8.76666666666667 0.86
8.77333333333333 0.87
9.04666666666667 0.88
9.07333333333333 0.89
9.25333333333333 0.9
9.94666666666667 0.91
10.24 0.92
10.5133333333333 0.93
10.82 0.94
11.42 0.95
11.6933333333333 0.96
11.9666666666667 0.97
12.0333333333333 0.98
12.0866666666667 0.99
14.74 1
};
\addlegendentry{F=4}
\addplot [semithick, color3, const plot mark left]
table {%
-inf 0
0.0133333333333333 0.0196078431372549
0.0333333333333333 0.0392156862745098
0.16 0.0588235294117647
0.4 0.0784313725490196
0.493333333333333 0.0980392156862745
0.5 0.117647058823529
0.593333333333333 0.137254901960784
0.6 0.156862745098039
0.633333333333333 0.176470588235294
0.826666666666667 0.196078431372549
0.846666666666667 0.215686274509804
0.886666666666667 0.235294117647059
0.933333333333333 0.254901960784314
1 0.274509803921569
1.02666666666667 0.294117647058824
1.14666666666667 0.313725490196078
1.19333333333333 0.333333333333333
1.3 0.352941176470588
1.40666666666667 0.372549019607843
1.48 0.392156862745098
1.65333333333333 0.411764705882353
1.66 0.431372549019608
1.68 0.450980392156863
1.7 0.470588235294118
1.74666666666667 0.490196078431373
1.76 0.509803921568627
1.92 0.529411764705882
1.94 0.549019607843137
1.94 0.568627450980392
1.95333333333333 0.588235294117647
1.97333333333333 0.607843137254902
2.02666666666667 0.627450980392157
2.19333333333333 0.647058823529412
2.34 0.666666666666667
2.46666666666667 0.686274509803922
2.48 0.705882352941177
2.66 0.725490196078431
2.73333333333333 0.745098039215686
2.88666666666667 0.764705882352941
2.96666666666667 0.784313725490196
3.08666666666667 0.803921568627451
3.09333333333333 0.823529411764706
3.55333333333333 0.843137254901961
3.94 0.862745098039216
4.00666666666667 0.882352941176471
4.1 0.901960784313726
4.35333333333333 0.92156862745098
4.44666666666667 0.941176470588235
4.49333333333333 0.96078431372549
4.88 0.980392156862745
5.32 1
};
\addlegendentry{Optimal}
\end{axis}

\end{tikzpicture}
    \caption{CDF of the percentage of users dropped per episodes.}
    \label{fig:disconnection}
\end{figure}

\subsection{Results}
We compare our HMARL implementation against static when $F=3$ or $F=4$, equivalent to 3GPP's split 7-2 and 7-1, for both users groups as common configurations in the \gls{oran} architecture. These baselines enable MC for users, while have presenting a good tradeoff between resource consumption and E2E delay. Additionally, results from an optimal solver for the problem are plotted. The optimal solver checks every combination of FS for each user group and \gls{cca}, deploying then the configuration that maximize the reward functions. 
\subsubsection{Constraint Violations}
In a constrained environment, the network needs to allocate resources so that maximum possible number of users get served with the required service continuity and data rate. \autoref{fig:viol} presents the ratio of timesteps per episode presenting violation for either the GOPS load constraint in \eqref{c1} or midhaul load constraint in \eqref{c2}. 
Results from \autoref{fig:viol} show that FS $3$ and FS $4$ do not impact the midhaul resources needed significantly, almost never incurring in violation of \eqref{c2}. 
This, however, is not true for the violation of \eqref{c1}. Static baselines, in fact, fail to comply with \gls{GOPS} constraints for $34\%$-$37\%$ of the time on average.
On the other hand, HMARL's ability to learn the required resources at given network state,  leads it to achieve lower violation ratios, incurring in GOPS constraint violations $13\%$ of the time and close to $1\%$ for the midhaul constraint violation. This leads to significantly fewer dropped users leading to less service disruption and higher reliability as seen in Figure~\ref{fig:disconnection}. When compared to the optimal policy, it is clear that \gls{hmarl}'s policies closely match the best possible policy.
\begin{figure}
    \centering
\begin{tikzpicture}

\definecolor{color0}{HTML}{004488}
\definecolor{color1}{HTML}{DDAA33}
\definecolor{color2}{HTML}{BB5566}
\definecolor{color3}{HTML}{000000}
\definecolor{color4}{HTML}{AAAAAA}

\begin{axis}[
legend cell align={left},
legend style={fill opacity=0.8, draw opacity=1, text opacity=1, draw=white!80!black, at={(0.00,0.5)},anchor=west},
tick align=outside,
tick pos=left,
width=0.97\linewidth,
height=0.5\linewidth,
x grid style={white!69.0196078431373!black},
xlabel={Ratio of non-transitional users (\%)},
xmajorgrids,
xmin=0.65, xmax=1.001,
xtick style={color=black},
xtick={0.6,0.65,0.7,0.75,0.8,0.85,0.9,0.95,1,1.05},
xticklabels={60,65,70,75,80,85,90,95,100},
y grid style={white!69.0196078431373!black},
ylabel={CDF},
ymajorgrids,
ymin=0, ymax=1,
ytick style={color=black},
ytick={0,0.2,0.4,0.6,0.8,1},
yticklabels={0.0,0.2,0.4,0.6,0.8,1.0},
font={\footnotesize},
]
\addplot [semithick, color0, const plot mark left]
table {%
-inf 0
0.766023241103863 0.02
0.776029409694769 0.04
0.787611146781275 0.06
0.793852722462887 0.08
0.800655033274849 0.1
0.800793303087747 0.12
0.809194227949789 0.14
0.810800976420094 0.16
0.815123827524991 0.18
0.81708721528978 0.2
0.81904361972247 0.22
0.819853777382744 0.24
0.821503410738954 0.26
0.822065071504211 0.28
0.82371559575055 0.3
0.82638521345217 0.32
0.828555876318052 0.34
0.832545677319696 0.36
0.836447533695646 0.38
0.838204651959311 0.4
0.841035014189838 0.42
0.845440070309807 0.44
0.848767634379438 0.46
0.849043338471372 0.48
0.851276252244521 0.5
0.852238736967189 0.52
0.853734849935292 0.54
0.863238647575127 0.56
0.864912232189062 0.58
0.868073214373558 0.6
0.870478209551954 0.62
0.874710634408957 0.64
0.878895981116519 0.66
0.880445443399016 0.68
0.883340495492395 0.7
0.891496663138539 0.72
0.89571389167385 0.74
0.895718211171116 0.76
0.897361357826003 0.78
0.900662730470723 0.8
0.904216848610327 0.82
0.909669754292943 0.84
0.910458610791544 0.86
0.912892440644401 0.88
0.914137511039685 0.9
0.926041328828247 0.92
0.926947354978553 0.94
0.950660337436386 0.96
0.960866053500938 0.98
0.96192337130127 1
};
\addlegendentry{HMARL}
\addplot [semithick, color1, const plot mark left]
table {%
-inf 0
0.688628828676302 0.0142857142857143
0.71330459250894 0.0285714285714286
0.716430642509253 0.0428571428571429
0.74346666834646 0.0571428571428571
0.743830497456575 0.0714285714285714
0.744971674337102 0.0857142857142857
0.747868403906019 0.1
0.751316385070184 0.114285714285714
0.75338443444884 0.128571428571429
0.757052806530249 0.142857142857143
0.762993745946849 0.157142857142857
0.764370168838769 0.171428571428571
0.764415811101336 0.185714285714286
0.769260936925327 0.2
0.774263220146431 0.214285714285714
0.775197260194948 0.228571428571429
0.775325104250864 0.242857142857143
0.776310212799038 0.257142857142857
0.781395199691202 0.271428571428571
0.785347170640897 0.285714285714286
0.785722740703938 0.3
0.78577493173946 0.314285714285714
0.786201488582264 0.328571428571429
0.789324340117207 0.342857142857143
0.789506335084889 0.357142857142857
0.790791052407299 0.371428571428571
0.791722259301988 0.385714285714286
0.796753629156814 0.4
0.801452383528032 0.414285714285714
0.803255217999213 0.428571428571429
0.805244243351435 0.442857142857143
0.805293182595843 0.457142857142857
0.806484007232429 0.471428571428571
0.806541190607991 0.485714285714286
0.807205627388492 0.5
0.807929589656735 0.514285714285714
0.810917077042093 0.528571428571429
0.812115002110651 0.542857142857143
0.812255906337895 0.557142857142857
0.814791729910449 0.571428571428571
0.815057312405587 0.585714285714286
0.815683077340915 0.6
0.816683101947589 0.614285714285714
0.817070477586556 0.628571428571429
0.818430035626718 0.642857142857143
0.81997013444645 0.657142857142857
0.822254290382231 0.671428571428571
0.822864912993302 0.685714285714286
0.826089622481412 0.7
0.826688508903016 0.714285714285714
0.826732867079894 0.728571428571429
0.82687431872306 0.742857142857143
0.831002103047813 0.757142857142857
0.833340197435152 0.771428571428571
0.834780043081771 0.785714285714286
0.840788935636735 0.8
0.841306818201076 0.814285714285714
0.841991226760242 0.828571428571429
0.842140110789891 0.842857142857143
0.845954134043627 0.857142857142857
0.84969991240556 0.871428571428571
0.851807255939514 0.885714285714286
0.853542722999461 0.9
0.862016177421511 0.914285714285714
0.867228755963786 0.928571428571429
0.879810408978964 0.942857142857143
0.881615061903326 0.957142857142857
0.884899742142446 0.971428571428571
0.887608266195039 0.985714285714286
0.894841732212613 1
};
\addlegendentry{F=3}
\addplot [semithick, color2, const plot mark left]
table {%
-inf 0
0.736788126135196 0.0196078431372549
0.74516199657689 0.0392156862745098
0.74814627790423 0.0588235294117647
0.74845814222636 0.0784313725490196
0.761749050104485 0.0980392156862745
0.762955738674382 0.117647058823529
0.76616013848819 0.137254901960784
0.770868010491783 0.156862745098039
0.776331433506962 0.176470588235294
0.778218449710727 0.196078431372549
0.780569625272744 0.215686274509804
0.781148056861925 0.235294117647059
0.784492728013151 0.254901960784314
0.786600220220349 0.274509803921569
0.786782562298018 0.294117647058824
0.787646452528318 0.313725490196078
0.790431662877911 0.333333333333333
0.793435427720992 0.352941176470588
0.794248227147691 0.372549019607843
0.794900636497299 0.392156862745098
0.796802928736791 0.411764705882353
0.803850040901833 0.431372549019608
0.804037997341545 0.450980392156863
0.805991381123609 0.470588235294118
0.808561261784037 0.490196078431373
0.811058899759605 0.509803921568627
0.813584203989374 0.529411764705882
0.814846558244813 0.549019607843137
0.815201258815781 0.568627450980392
0.815268031804645 0.588235294117647
0.817862533746534 0.607843137254902
0.819144756453486 0.627450980392157
0.820391267548876 0.647058823529412
0.820442559287348 0.666666666666667
0.821163166868622 0.686274509803922
0.823382785913162 0.705882352941177
0.823662640416972 0.725490196078431
0.82485091616178 0.745098039215686
0.826365533087558 0.764705882352941
0.826444508682977 0.784313725490196
0.833744109746367 0.803921568627451
0.835622367946993 0.823529411764706
0.836486397153007 0.843137254901961
0.836803276507796 0.862745098039216
0.838264742961296 0.882352941176471
0.841444433066439 0.901960784313726
0.843164063423369 0.92156862745098
0.852120985165074 0.941176470588235
0.858569848162545 0.96078431372549
0.859144324080728 0.980392156862745
0.875560444977093 1
};
\addlegendentry{F=4}
\addplot [semithick, color3, const plot mark left]
table {%
-inf 0
0.745391356361479 0.0196078431372549
0.785628052002908 0.0392156862745098
0.788100149664814 0.0588235294117647
0.790479578305286 0.0784313725490196
0.805108036203214 0.0980392156862745
0.818272355906079 0.117647058823529
0.826201248459245 0.137254901960784
0.833044189801074 0.156862745098039
0.838121973159238 0.176470588235294
0.84281468147676 0.196078431372549
0.844749111335877 0.215686274509804
0.84483470939282 0.235294117647059
0.850921725328136 0.254901960784314
0.851018808427193 0.274509803921569
0.853043006392568 0.294117647058824
0.854584007344119 0.313725490196078
0.859730996498769 0.333333333333333
0.864331761816102 0.352941176470588
0.871340314945631 0.372549019607843
0.873792826340382 0.392156862745098
0.881063543162914 0.411764705882353
0.881118420178617 0.431372549019608
0.882137925100599 0.450980392156863
0.882277055285351 0.470588235294118
0.886428619907287 0.490196078431373
0.886555578294632 0.509803921568627
0.893007436868722 0.529411764705882
0.893263493332811 0.549019607843137
0.893498146421621 0.568627450980392
0.896215983527226 0.588235294117647
0.897632080524065 0.607843137254902
0.899920590447454 0.627450980392157
0.907056058021561 0.647058823529412
0.908706477812419 0.666666666666667
0.910547519652706 0.686274509803922
0.914823464880883 0.705882352941177
0.916019641392759 0.725490196078431
0.91841915001715 0.745098039215686
0.927814360240439 0.764705882352941
0.931023523759908 0.784313725490196
0.93185906268986 0.803921568627451
0.933748893866009 0.823529411764706
0.936149494813101 0.843137254901961
0.938305011672672 0.862745098039216
0.948201289908213 0.882352941176471
0.951870786701221 0.901960784313726
0.956666286436252 0.92156862745098
0.958017470587722 0.941176470588235
0.95833305508233 0.96078431372549
0.962446143772454 0.980392156862745
0.978750915750916 1
};
\addlegendentry{Optimal}
\end{axis}

\end{tikzpicture}
    \caption{CDF of non-transitional users maintaining service continuity.}
    \label{fig:non-transitional}
\end{figure}
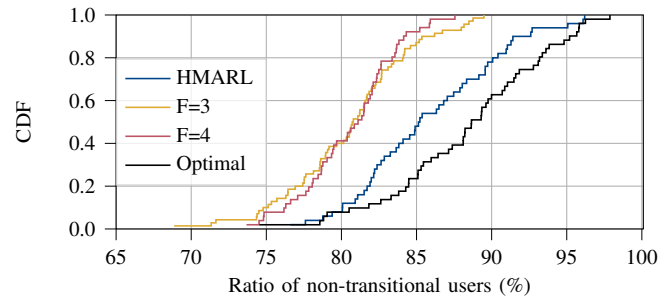
%
\subsubsection{Service continuity}
By handling the network resources better than the other baselines, HMARL can reduce the amount of users dropped due to the constraint violation. This is clearly reflected on the amount of non-transitional users able to achieve and maintain service continuity. 
Figure~\ref{fig:non-transitional} shows the cumulative distribution function (CDF) of the ratio of non-transitional users maintaining service continuity for the episode duration.
FSs $3$ and $4$ are experimentally the only two FSs for transitional users able to satisfy users' delay requirements. In particular, FS $4$ is able to grant a significant performance improvement with respect to FS $3$, at the cost of an increased computational cost in terms of \gls{GOPS}. For this reason, the performance of the two baselines are comparable. FS $3$ increases the delay perceived by users  but is lighter on \gls{GOPS}. On the other hand, FS $4$ grants lower latency, at the cost of higher user dropping rate. HMARL is able to capture this behavior, learning how to  use FS $3$ when possible, leaving the EC agents with more resources to use, and FS $4$ when available resources are enough to support it. By doing so, EC agents can satisfy the needs of more users, leading to a visibly higher amount of users achieving service continuity inside their CCA. The policies learnt are able to increase the amount of supported users of $5\%$ to $10\%$ with respect to the static configurations. As for the previous tests, \gls{hmarl}'s results are close to the optimal policy's results.
\subsubsection{Policy scalability}
\begin{figure}
    \centering
\begin{tikzpicture}

\definecolor{color0}{HTML}{004488}
\definecolor{color1}{HTML}{DDAA33}
\definecolor{color2}{HTML}{BB5566}
\definecolor{color3}{HTML}{000000}
\definecolor{color4}{HTML}{AAAAAA}

\begin{axis}[
legend cell align={left},
legend style={
  fill opacity=0.8,
  draw opacity=1,
  text opacity=1,
  at={(1,0.25)},
  anchor=east,
  draw=white!80!black,
},
width=\linewidth,
height=0.6\linewidth,
tick align=outside,
tick pos=left,
x grid style={white!69.0196078431373!black},
xlabel={GOPS available at every EC ($G_{th}$)},
xmajorgrids,
xmin=1795, xmax=1905,
xtick={1800,1825,1850,1875,1900},
xticklabels={$14k$,$15k$,$16k$,$17k$,$18k$},
xtick style={color=black},
y grid style={white!69.0196078431373!black},
ylabel={Ratio of users (\%)},
ymajorgrids,
ymin=0.715877691331785, ymax=0.95,
ytick={0.75,0.80,0.85,0.90,0.95},
yticklabels={75,80,85,90},
ytick style={color=black},
font={\footnotesize},
]
\addplot [semithick, color0, mark=*, mark size=3, mark options={solid}]
table {%
1800 0.784230604561542
1825 0.806592615759745
1850 0.851757494605855
1875 0.89215305187686
1900 0.900997285731897
};
\addlegendentry{HMARL}
\addplot [semithick, color1, mark=*, mark size=3, mark options={solid}]
table {%
1800 0.806617475615462
1825 0.811973710397707
1850 0.807567608522614
1875 0.854379787714104
1900 0.859921562162581
};
\addlegendentry{F=$3$}
\addplot [semithick, color2, mark=*, mark size=3, mark options={solid}]
table {%
1800 0.730863270667943
1825 0.724692910112742
1850 0.811058899759605
1875 0.854592163239549
1900 0.872982693860269
};
\addlegendentry{F=$4$}
\addplot [semithick, color3, mark=*, mark size=3, mark options={solid}]
table {%
1800 0.849168459556006
1825 0.84203056301744
1850 0.877131257182077
1875 0.926698109246169
1900 0.935171150037608
};
\addlegendentry{Optimal}
\end{axis}

\end{tikzpicture}
    \caption{Ratio of users achieving service continuity with respect to the amount of GOPS available at every EC.}
    \label{fig:scalability}
\end{figure}
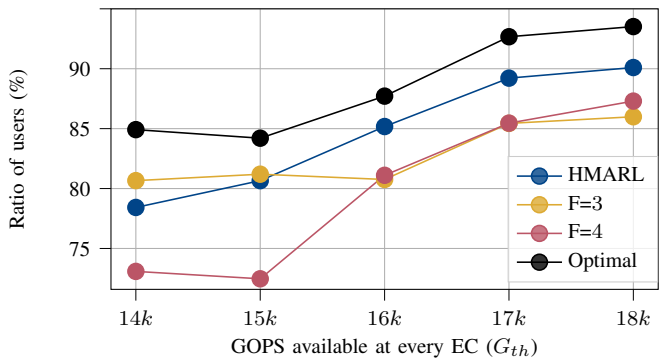

Once trained, HMARL is able to understand the resources usage in the network and deliver the close to optimal functional split configuration to the two users group. To examine the generalization capabilities of the trained policies, we applied them in unseen network scenarios to investigate its adaptability and generalization capabilities. In Figure \ref{fig:scalability}, the ratio of users with service continuity with respect to a varying $G_{th}$ is plotted. As noticeable, the policies learnt by HMARL remain effective as resources become abundant: HMARL is able to exploit the increased resources to satisfy even more users than in the scenario it has been trained on. On the other hand, when decreasing the resources available, the policies learnt by HMARL slightly increases the number of users dropped, leading to a performance loss with respect to static baselines.

\section{Conclusion}\label{outlook}
In this paper, we addressed the problem of maintaining service continuity and enabling soft handovers for transitional users without compromising the performance of other users in the network. Our proposed solution, the HMARL algorithm, demonstrated significant benefits for all users, particularly non-transitional ones. By learning adaptive policies, HMARL not only reduces the number of dropped users but also allocates sufficient resources to meet their service requirements. Even in previously unseen scenarios, the algorithm effectively utilizes available resources. However, HMARL has its limitations: training the model for real-world deployment would require substantial time, and the complexity of the infrastructure makes it prone to the curse of dimensionality. In particular, the training complexity of such an infrastructure scales linearly with the number of agents and parameters, making it a major challenge  in a practical O-RAN system.  We therefore aim to address these constraints to make HMARL a practical solution for practical O-RAN deployments.

\section*{Acknowledgement}

This work was supported in part by the CELTIC-NEXT Project, 6G for Connected Sky (6G-SKY), with funding received from the Vinnova, Swedish Innovation Agency.

\bibliographystyle{IEEEtran}
\bibliography{lib.bib}

\begin{thebibliography}{10}
\providecommand{\url}[1]{#1}
\csname url@samestyle\endcsname
\providecommand{\newblock}{\relax}
\providecommand{\bibinfo}[2]{#2}
\providecommand{\BIBentrySTDinterwordspacing}{\spaceskip=0pt\relax}
\providecommand{\BIBentryALTinterwordstretchfactor}{4}
\providecommand{\BIBentryALTinterwordspacing}{\spaceskip=\fontdimen2\font plus
\BIBentryALTinterwordstretchfactor\fontdimen3\font minus \fontdimen4\font\relax}
\providecommand{\BIBforeignlanguage}[2]{{%
\expandafter\ifx\csname l@#1\endcsname\relax
\typeout{** WARNING: IEEEtran.bst: No hyphenation pattern has been}%
\typeout{** loaded for the language `#1'. Using the pattern for}%
\typeout{** the default language instead.}%
\else
\language=\csname l@#1\endcsname
\fi
#2}}
\providecommand{\BIBdecl}{\relax}
\BIBdecl

\bibitem{mahmoud20216g}
H.~H.~H. Mahmoud \emph{et~al.}, ``{6G: A comprehensive survey on technologies, applications, challenges, and research problems},'' \emph{Trans. on Emerging Telecommunications Technologies}, vol.~32, no.~4, pp. 1--14, 2021.

\bibitem{rihan2023ran}
M.~Rihan \emph{et~al.}, ``{RAN Functional Split Options for Integrated Terrestrial and Non-Terrestrial 6G Networks},'' in \emph{IEEE International Japan-Africa Conference on Electronics, Communications, and Computations (JAC-ECC)}, 2023, pp. 152--158.

\bibitem{larsen_survey_2018}
L.~M. Larsen, A.~Checko, and H.~L. Christiansen, ``A survey of the functional splits proposed for {5G} mobile crosshaul networks,'' \emph{IEEE Commun. Surv. \& Tut.}, vol.~21, no.~1, pp. 146--172, 2018.

\bibitem{bassoy_coordinated_2017}
S.~Bassoy \emph{et~al.}, ``Coordinated multi-point clustering schemes: {A} survey,'' \emph{IEEE Comm. Surveys \& Tutorials}, vol.~19, no.~2, pp. 743--764, 2017.

\bibitem{oran}
M.~Polese \emph{et~al.}, ``Understanding o-ran: Architecture, interfaces, algorithms, security, and research challenges,'' \emph{IEEE Communications Surveys \& Tutorials}, vol.~25, no.~2, pp. 1376--1411, 2023.

\bibitem{Meer2024}
I.~A. Meer \emph{et~al.}, ``{Learning Based Dynamic Cluster Reconfiguration for UAV Mobility Management with 3D Beamforming},'' in \emph{Proc. IEEE ICMLCN}, May 2024, p. 486–491.

\bibitem{meer2024hierarchical}
------, ``{Hierarchical Multi-Agent DRL Based Dynamic Cluster Reconfiguration for UAV Mobility Management},'' \emph{arXiv preprint arXiv:2412.16167}, 2024.

\bibitem{wang_handover_2016}
X.~Wang \emph{et~al.}, ``Handover reduction in virtualized cloud radio access networks using {TWDM}-{PON} fronthaul,'' \emph{J. of Opt. Commun. and Netw.}, vol.~8, no.~12, pp. B124--B134, 2016.

\bibitem{alba_dynamic_2022}
A.~M. Alba and W.~Kellerer, ``Dynamic {Functional} {Split} {Adaptation} in {Next}-{Generation} {Radio} {Access} {Networks},'' \emph{IEEE Trans. on Network and Service Management}, vol.~19, no.~3, pp. 3239--3263, 2022.

\bibitem{matoussi_user-centric_2021}
\BIBentryALTinterwordspacing
S.~Matoussi, ``User-{Centric} {Slicing} with {Functional} {Splits} in {5G} {Cloud}-{RAN},'' Ph.D. dissertation, Sorbonne Université, Jan. 2021. [Online]. Available: \url{https://theses.hal.science/tel-03951250}
\BIBentrySTDinterwordspacing

\bibitem{alabbasi_delay-aware_2017}
A.~Alabbasi and C.~Cavdar, ``Delay-aware green hybrid {CRAN},'' in \emph{Proc. IEEE WiOpt}, 2017, pp. 1--7.

\bibitem{wang_interplay_2017}
X.~Wang \emph{et~al.}, ``Interplay of energy and bandwidth consumption in {CRAN} with optimal function split,'' in \emph{Proc. IEEE ICC}, 2017, pp. 1--6.

\bibitem{noauthor_small_nodate}
\BIBentryALTinterwordspacing
``Small {Cell} {Forum} {Releases}.'' [Online]. Available: \url{https://scf.io/en/documents/159\_-\_Small\_cell\_virtualization\_functional\_splits\_and\_use\_cases.php}
\BIBentrySTDinterwordspacing

\bibitem{masoudi_energy-optimal_2022}
M.~Masoudi, O.~T. Demir, J.~Zander, and C.~Cavdar, ``Energy-{Optimal} {End}-to-{End} {Network} {Slicing} in {Cloud}-{Based} {Architecture},'' \emph{IEEE Open Journal of the Communications Society}, vol.~3, pp. 574--592, 2022.

\bibitem{wang_centralize_2017}
X.~Wang \emph{et~al.}, ``Centralize or distribute? {A} techno-economic study to design a low-cost cloud radio access network,'' in \emph{IEEE ICC}, 2017, pp. 1--7.

\bibitem{salehi_reliability_2023}
F.~Salehi, M.~Ozger, and C.~Cavdar, ``Reliability and delay analysis of 3-dimensional networks with multi-connectivity: Satellite, haps, and cellular communications,'' \emph{IEEE Transactions on Network and Service Management}, vol.~21, no.~1, pp. 437--450, 2024.

\bibitem{busoniu_multi-agent_2010}
L.~Buşoniu, R.~Babuška, and B.~De~Schutter, ``Multi-agent {Reinforcement} {Learning}: {An} {Overview},'' in \emph{Innovations in {Multi}-{Agent} {Systems} and {Applications} - 1}.\hskip 1em plus 0.5em minus 0.4em\relax Berlin, Heidelberg: Springer, 2010, pp. 183--221.

\bibitem{brockman2016openai}
G.~Brockman \emph{et~al.}, ``Openai gym,'' \emph{arXiv preprint arXiv:1606.01540}, 2016.

\bibitem{liang2018rllib}
E.~Liang \emph{et~al.}, ``{RLlib: Abstractions for distributed reinforcement learning},'' in \emph{Proc. Int. Conf. Mach. Learn}, 2018, pp. 3053--3062.

\bibitem{schulman_proximal_2017}
J.~Schulman, F.~Wolski, P.~Dhariwal, A.~Radford, and O.~Klimov, ``Proximal {Policy} {Optimization} {Algorithms},'' Aug. 2017, arXiv:1707.06347.

\end{thebibliography}
\end{document}